\documentclass[11pt,a4paper]{article}
\pdfoutput=1
\usepackage{jcappub}
\usepackage{slashed}

\def\be{\begin{equation}}
\def\ee{\end{equation}}
\def\bea{\begin{eqnarray}}
\def\eea{\end{eqnarray}}

\begin{document}

\title{Dual Field Theories of Quantum Computation}

\author{Vitaly Vanchurin}

\emailAdd{vvanchur@d.umn.edu}

\date{\today}

\affiliation{Department of Physics and Astronomy, University of Minnesota, Duluth, Minnesota, 55812}

\abstract{ 
Given two quantum states of $N$ q-bits we are interested to find the shortest quantum circuit consisting of only one- and two- q-bit gates that would transfer one state into another. We call it the quantum maze problem for the reasons described in the paper. We argue that in a large $N$ limit the quantum maze problem is equivalent to the problem of finding a semiclassical trajectory of some lattice field theory (the dual theory) on an $N+1$ dimensional space-time with geometrically flat, but topologically compact spatial slices. The spatial fundamental domain is an $N$ dimensional hyper-rhombohedron, and the temporal direction describes transitions from an arbitrary initial state to an arbitrary target state and so the initial and final dual field theory conditions are described by these two quantum computational states. We first consider a complex Klein-Gordon field theory and argue that it can only be used to study the shortest quantum circuits which do not involve generators composed of tensor products of multiple Pauli $Z$ matrices.  Since such situation is not generic we call it the $Z$-problem.  On the dual field theory side the $Z$-problem corresponds to massless excitations of the phase (Goldstone modes) that we attempt to fix using Higgs mechanism. The simplest dual theory  which does not suffer from the massless excitation (or from the $Z$-problem) is the Abelian-Higgs model which we argue can be used for finding the shortest quantum circuits. Since every trajectory of the field theory is mapped directly to a quantum circuit,  the shortest quantum circuits are identified with semiclassical trajectories. We also discuss the complexity of an actual algorithm that uses a dual theory prospective for solving the quantum maze problem and compare it with a geometric approach. We argue that it might be possible to solve the problem in sub-exponential time in $2^N$, but for that we must consider the Klein-Gordon theory on curved spatial geometry and/or more complicated (than $N$-torus) topology. }

\maketitle

\section{Introduction}

Consider a quantum system of $N$ q-bits whose states can be described by unit vectors in $2^N$-complex dimensional  Hilbert space. The unit size of the sphere indicates that all of the points are within distance of $O(1)$ from each other if you were allowed to move along geodesics on the unit sphere. Now imagine that you are only allowed to move in $O(N^2)$ orthogonal directions out of $O(2^{N})$. More precisely, at any point you are allowed to only apply $O(N)$ of one- q-bit gates or $O(N^2)$ of two- q-bit gates.  Then the relevant question is: what is the shortest distance connecting an arbitrary pair of points on the unit sphere? This is like playing a very high-dimensional maze with a lot of walls and very few pathways.

There are two motivations to study the ``quantum maze'': one computational and one physical.  First of all if we knew how to solve the ``quantum maze'' problem we would be able to design the most efficient quantum algorithms or in other words to construct the shortest quantum circuits that can transform some simple initial state to the desired target state. A problem which is known to be double exponentially hard $O\left (2^{2^N} \right )$ (See Ref. \cite{book} for a pedagogical discussion of computational complexities in context of quantum information theory). The second reason has its roots in black-hole physics. It was conjectured that black-holes are the fastest quantum computers in nature \cite{Brown:2015bva} and so in some sense the black-holes know how to solve the ``quantum maze''. Some other recent applications of the theory of computational complexities in context of black-hole physics were discussed in Ref. \cite{Harlow} and in Refs. \cite{Susskind1, Susskind2} in an attempt to tackle the firewall problem \cite{Firewall}. 

In addition, a large body of work is directed towards establishing connections between special kinds of quantum circuits (known as tensor networks) in context of the AdS/CFT correspondence \cite{Swingle, Dong, Pastawski, Czech}. Indeed the tensor networks provide an interesting new prospective on how the spacetime (on the AdS side) might emerge from a holographic state (on the CFT side) by identifying the spacetime with a tensor network which would produce such holographic state. In Refs.  \cite{Brown:2015bva, Brown:2015lvg} the authors went even further and first conjectured and then verified that the complexity of the holographic CFT states are dual to the actions over certain patches in the AdS spacetime.  

In this paper we are going to expand and build upon the action-complexity conjecture \cite{Brown:2015bva}, but we shall not be concerned with transitions from only simple states to only holographic states. Instead we will study transitions from an arbitrary initial  $|\psi_{\text{in}} \rangle $ to an arbitrary final  $|\psi_{\text{out}}\rangle $ state, i.e. the quantum maze problem. Nevertheless we will still conjecture that there must exist a (yet to be discovered) dual field theory whose Euclidean action describes the computational complexity of the smallest quantum circuit connecting the two states, i.e. 
\be
{\cal C} (|\psi_{\text{out}}\rangle, |\psi_{\text{in}}\rangle)  =  S_E[\Phi] \label{eq:conjecture1}
\ee
where $\Phi$ is a collective notation for all fields. Note that both sides of Eq. \eqref{eq:conjecture1} are not uniquely specified: the left hand side depends on  size of Hilbert space and on the collection of fundamental gates and the right hand side depends on the Lagrangian and on the region of integration, but it is quite possible that for certain collections of fundamental gates there exist a Lagrangian with desired properties. 

The paper is organized as follows. In Sec. \ref{Sec:PathIntegral} we define path integrals for the dual theories of quantum computation and discuss the symmetries that the dual theory should possess. In Sec. \ref{Sec:FieldTheory} we argue that the dual theory can be conveniently described as a Klein-Gordon lattice field theory on an $N+1$ dimensional space-time with geometrically flat, but topologically compact spatial slices. In Sec. \ref{Sec:QuantumCircuits} we show how one can map classical solutions on the dual theory side to quantum circuits on the quantum computation side. In Sec. \ref{Sec:AbelianHiggs} we identify the $Z$-problem of the Klein-Gordon theory which corresponds to the massless excitations problem (Goldstone mode) and attempt to fix it using Higgs mechanism. In Sec. \ref{Sec:QuantumMaze} we discuss how the dual theory description can be used for solving the ``quantum maze'' problem in sub-exponential time, but for that we must consider curved spatial geometry and/or more complicated spatial topologies. In Sec. \ref{Sec:Summary} we summaries the main results of the paper.

\section{Path Integrals for Dual Theories}\label{Sec:PathIntegral}

Our initial task is to describe a possible construction of the dual field theories for a system of $N$ q-bits with fundamental gates consisting of all one- q-bit and two q-bit operations.  Since the total number of degrees of freedom of such system is finite (dimensionality of Hilbert space is only $2^N$),  we are directed towards  dual theories on the lattice with finite fundamental domain, i.e.
  \be
{\cal C}  (|\psi_{\text{out}}\rangle, |\psi_{\text{in}}\rangle)  = \int_0^T dt_E L_E(\Phi^i, \dot{\Phi}^i)
\ee
 where the index $i$ enumerates the lattice points. There are many choices on how the initial and final quantum computational states can manifest themselves in the dual theory, but in the spirit of the holographic ideas we are going to make the following identification
\bea
| \psi_{\text{in}} \rangle = F^i(\Phi^j(0)) |i\rangle \\ 
| \psi_{\text{out}} \rangle = F^i(\Phi^j(T)) |i\rangle \label{eq:boundary_conditions}
\eea
where $i=\{0,1\}^N$ is an integer in base two and the Einstein summation convention over repeated indices is implied. In other words the initial and final quantum states determine the initial and final configurations of fields in the dual theory according to the function $F$ which is yet to be specified. 

For simplicity we first assume that the theory contains only $2^N$ complex degrees of freedom $\varphi^i$ with boundary conditions set as
\bea
| \psi_{\text{in}} \rangle = \varphi^i(0) |i\rangle \notag\\ 
| \psi_{\text{out}} \rangle = \varphi^i(T) |i\rangle, \label{eq:boundary_conditions2}
\eea
where, once again, the summation over repeated indices is implied. In the limit of small $\hbar$ only a single ``classical'' trajectory $\varphi_{cl}$ would contribute to the overall path integral whose Euclidean action was conjectured to give the quantum computational complexity.\footnote{Note that $\hbar$ is a constant which will appear in the partition function for our dual theory \eqref{eq:partition}, but it may or may not be related to the Planck constant. } This suggests the following expression for complexity in term of the (yet to be defined) partition function
\bea
{\cal C}  (|\psi_{\text{out}}\rangle, |\psi_{\text{in}}\rangle)  &=&   \int_0^T dt_E   L_E(\varphi_{cl}^i, \dot{\varphi}_{cl}^i) \label{eq:Euclidean_complexity} \\ &=&  \lim_{\hbar\rightarrow0 } \hbar^2 \frac{d}{d\hbar} \log\left({{\cal Z}_\hbar(|\psi_{\text{out}}\rangle, |\psi_{\text{in}}\rangle)} \right).
\eea
On the other hand if $\hbar$ has a physical significance (for example describes how well the corresponding quantum computer is isolated, or how precise we wish to approximate the quantum evolution to the target state), then it might be more appropriate to describe the difficulty of certain computational task with what we can call the computational free energy, 
\be
{\cal F}  (|\psi_{\text{out}}\rangle, |\psi_{\text{in}}\rangle)  = - \hbar \log\left({{\cal Z}_\hbar(|\psi_{\text{out}}\rangle, |\psi_{\text{in}}\rangle)} \right)
\ee
which contains information about all of the trajectories including, but not limited to only,  ``classical'' trajectories. 

Now that the connection between partition function and complexity was established, the next step is to construct the dual theory Lagrangian. Our task will be to find a Lorentzian dual theory which is connected to the Euclidean dual theory through Wick rotation of the time coordinate $t=-i t_E$. From the boundary conditions in \eqref{eq:boundary_conditions2} it is clear that we are dealing with $2^N$ complex degrees of freedom $\varphi^i$ which represent evolution of our state vector from initial  $|\psi_{\text{in}}\rangle$ to final $|\psi_{\text{out}}\rangle$  state. This imposes a constraint that all trajectories must remain normalized throughout the evolution,
\be
\varphi_i(t) \varphi^i(t) =1
\ee
where 
\be \varphi^*_i \equiv \varphi^{i}. \ee 
In addition we shall demand that the Lagrangian is invariant under $U(2^N)$ transformation, i.e.
\be
\varphi^i \rightarrow U^{i}_{\phantom{i}j} \varphi^j, 
\ee
and also invariant under arbitrary  permutations of bits. The former property guarantees that complexities of trajectories are invariant if the initial and final states are transformed simultaneously.  For the latter property it will be convenient to introduce a function $h(i,j)$ which measures, the so-called, Hamming distance between classical strings of bits representing $i$  and $j$, i.e. the number of positions in which corresponding values of bits are different. For example, $h(0,7)=h(000_2,111_2)= 3,  h(2, 6) =h(010_2, 110_2) = 1, h(3,3)=h(011_2,011_2)=0$, etc. 

By combining all three conditions the leading terms of the Lagrangian can be written as
\be
L(\varphi_i, \dot{\varphi}_i) = A \frac{1}{2} \dot{\varphi}_i\dot{\varphi}^i + \lambda (\varphi_i \varphi^i  - 1)+ f(h(i,j)) \varphi_i \varphi^j +... \label{eq:discrete} 
\ee
where $f(h)$ is some function of Hamming distance. (Roughly speaking we expect the function to vanish if Hamming distance is larger than $2$, i.e. penalizing more than three q-bit gates.) To make the above expression covariant we can replace  $f(h(i, j))$ with a tensor defined as  $f^i_{\phantom{i}j} \equiv f(h(i, j))$ only in computational basis and to transform to other basis it must be treated as a rank $(1,1)$ tensor under $U\left(2^N\right)$ transformations. Thus we arrive at a Lorentzian path integral expression
\be
{\cal Z}(|\psi_{\text{out}}\rangle, |\psi_{\text{in}}\rangle) = \int^{| \psi_{\text{out}} \rangle = \varphi^i(T) |i\rangle }_{| \psi_{\text{in}} \rangle = \varphi^i(0) |i\rangle } d^{2^N} \varphi \;\; e^{\frac{i}{\hbar} \int_0^T dt \left (A \frac{1}{2} \dot{\varphi}_i\dot{\varphi}^i + \lambda (\varphi_i \varphi^i  - 1)+  f^i_{\phantom{i}j} \varphi_i \varphi^j  \right )} + ... \label{eq:partition}
\ee
(Note that the constraint $\lambda (\varphi_i\varphi^i -1)$ must be imposed before the path integral is evaluated.) If we put aside the constraining term (as it is not likely to be very significant in the large $N$ limit) we can improve the path integral representation by treating $\varphi$ as a field in $N+1$ dimensional space-time. 

\section{Klein-Gordon Dual Theory} \label{Sec:FieldTheory} 

\begin{figure}[]
\begin{center}
\includegraphics[width=0.5\textwidth]{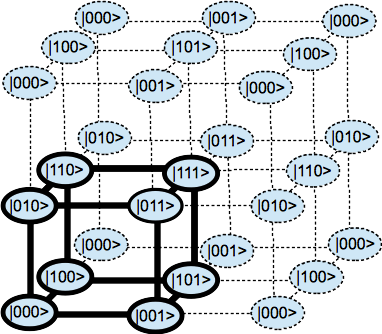}
\caption{Spatial base manifold for a dual field theory of a quantum system of three q-bits. Lattice points are marked with respective computational basis vectors  and a single copy of the  Hamming cube is highlighted with solid bold lines.  \label{plot}}
\end{center}
\end{figure}

In this section we will argue that the path integral in \eqref{eq:partition} can be written as a quantum field theory path integral on $N$ dimensional torus with only $2^N$ lattice points, i.e. only two lattice points along each dimension of total length $2l$. (To illustrate the main idea, on Fig. \ref{plot} we plot the torus for $N=3$ q-bits where all lattice points are marked with respective computational basis vectors and opposite sides are assumed to be identified. The Hamming cube (highlighted with solid bold lines) represents a desired connectivity between lattice sites which is constructed in such a way  that the transformations involving a large number of q-bits are penalized.)
Then in a continuum limit the path integral would be given by 
\be
{\cal Z}(|\psi_{\text{out}}\rangle, |\psi_{\text{in}}\rangle)  = \int^{| \psi_{\text{out}} \rangle = \varphi^i(T) |i\rangle }_{| \psi_{\text{in}} \rangle = \varphi^i(0) |i\rangle } D\varphi e^{\frac{i}{\hbar} \int_0^T dt \int d^N x {\cal L}(\varphi(x), \partial_\mu\varphi(x)) } \label{eq:continuum}
\ee
where $\mu$ index corresponds to temporal direction as well as spatial directions.\footnote{We adopt a notation of using Latin indices  from the beginning of alphabet (e.g. $a,b,...$) for only spatial directions and Greek indices  (e.g. $\mu,\nu, ...$) for both temporal and spatial directions.} In what follows we will be interested in semiclassical (or nearly classical paths) as they would describe evolutions of state vectors in a quantum circuit with a finite (as opposed to infinite) number of operations of gates.

The reader might object that we are now considering a continuum limit of a lattice field theory where the lattice size has a physical meaning. In such a construction taking a limit of zero lattice distance is meaningless, but that is not what we attempt to do. In fact, despite of writing a continuum representation \eqref{eq:continuum} we shall only be interested in a very coarse-grained description of the path integral with only two lattice points along every dimension. However, the advantage of going from a discrete to continuum is that we can try to explicitly construct a local Lagrangian density which would suppress evolutions corresponding to multiple-q-bit gates and would enable us to understand the quantum maze problem from a dual field theory prospective.

\subsection{Dual Lattice Field Theory}

All of the interactions in our discrete Lagrangian \eqref{eq:discrete}  (i.e. $f(h(i,j) \left (\varphi_i \varphi^j + \varphi_j \varphi^i \right ) $) can be written by summing over self-interactions (Hamming distance 0),  nearest neighbors interactions (Hamming distance 1) and neighbor-of-neighbor interactions (Hamming distance 2).  If we assume that $f(h) = 0$ for $h>2$ then it is convenient  to denote the three relevant constants as $B \equiv f(0)$, $C \equiv f(1)$ and $D \equiv f(2)$.  Then $f(h(i,j) \left (\varphi_i \varphi^j + \varphi_j \varphi^i \right ) $  can be rewritten as one self-interaction term per  lattice site $i$
\be
B  \varphi_i\varphi^i ,
\ee
$N$ nearest neighbor interaction terms per lattice site $i$
\be
\frac{C}{2} (\varphi_i\varphi^j + \varphi_j\varphi^i) = - \frac{C}{2}   (\varphi_i-\varphi_j) (\varphi^i-\varphi^j) + \frac{C}{2}  \varphi_i \varphi^i + \frac{C}{2}  \varphi_j \varphi^j  
\ee
and $N (N-1)/2$  neighbor-of-neighbor interaction terms per  lattice site $i$
\bea 
\nonumber  \frac{D}{2}  ( \varphi_i\varphi^j + \varphi_j \varphi^i ) = & - & \frac{D}{2}  \left ( (\varphi_i - \varphi_k)(\varphi^i - \varphi^k) +(\varphi_k-\varphi_j)(\varphi^k-\varphi^j) \right ) \\
\nonumber  & - &\frac{D}{2}  \left (  (\varphi_i - \varphi_k) (\varphi^k - \varphi^j)+ (\varphi_k - \varphi_j) (\varphi^i - \varphi^k) \right ) \\
 & + & \frac{D}{2}  \left ( \varphi_i \varphi^i + \varphi_j \varphi^j \right )  \;\;\;\;
\eea
 where $k$ site is a neighbor to both $i$ and $j$ sites.

 In a continuum limit we get a mass term
\be
l^{-N-1} \left (B + N C + \frac{N (N-1)}{2} D \right ) \varphi_i\varphi^i \;\;\; = \;\;\; \frac{1}{2} m^2 \varphi^2, 
\ee
diagonal gradient terms
\bea
l^{1-N} \frac{A}{2}  \left (\frac{\varphi_i(t)-\varphi_i(t+l)}{l} \right)  \left (\frac{\varphi^i(t)-\varphi^i(t+l)}{l} \right)  = \frac{1}{2} g^{00} \partial_0 \varphi^* \partial_0 \varphi \\
l^{1-N} \left (  - \frac{C}{2}  -  (N-1) \frac{D}{2}   \right ) \left(\frac{\varphi_i-\varphi_j}{l}\right ) \left(\frac{\varphi^i-\varphi^j}{l}\right ) = \frac{1}{2} g^{aa} \partial_a \varphi^* \partial_a \varphi
\eea
and off-diagonal gradient terms
\be
l^{1-N} \frac{D}{2}  \left(\frac{\varphi_i-\varphi_k}{l} \right ) \left ( \frac{\varphi^k-\varphi^j}{l}\right )  = \frac{1}{2} g^{ab}  \partial_a \varphi^* \partial_b \varphi
\ee
where there are no summations over repeated indices on either side of the equations. We can now rewrite the quantum field theory Lagrangian as 
\be
 {\cal L}(\varphi(x), \partial_\mu\varphi(x)) = \frac{1}{2}  g^{\mu\nu} \partial_\mu \varphi^* \partial_\nu \varphi - \frac{1}{2} m^2 \varphi^* \varphi  \label{eq:Lagrangian}
\ee
where the``mass''  
\be
m^2 \equiv  - \left (2 B + 2 N C + N (N-1) D \right ) l^{-N-1} \label{eq:mass}
\ee
and the inverse ``metric''  is
\bea
g^{00}&  \equiv & A l^{1-N} \\
g^{aa}& \equiv  & - \left (C + (N-1) D \right ) l^{1-N}\\
g^{ab}&  \equiv  &  D l^{1-N}, \label{eq:inverse_metric}
\eea
where $a \neq b$. One can also invert the above expression to obtain the metric tensor
\bea
g_{00}& = &  \frac{ 1}{A} l^{N-1} \\
g_{aa}& = &  - \frac{  \left (C +  D \right ) }{ C( C+N D)} l^{N-1}\\
g_{ab}& = &  - \frac{ D}{ C(C+N D)}  l^{N-1}. \label{eq:metric}
\eea

Note that $m$ and $g_{\mu\nu}$ now contain all of the informations about computational complexities, but for our path integral to remain finite we need the mass squared and all but one eigenvalues of the metric to be negative which implies\footnote{Note that we use a ``mostly negative'' metric convention.}
\bea
A & > &0 \\
C & >  & 0 \\
D & >  & - \frac{C}{N} \\
B & <  & - NC - \frac{N (N-1)}{2} D.
\eea
Later in the paper we will invert the last inequality in order to reproduce tachyonic mass of the Abelian-Higgs model, but the path integral will remain finite due to the high order terms that we omitted here. 

The metric  \eqref{eq:metric} is geometrically flat and can be put into a form of Minkowski metric for arbitrary $N$. For example, in two spatial dimensions (i.e. for a system of two q-bits) we can preform the following coordinate transformation: 
\bea
x_{0} & = & x'_0 \sqrt{\frac{2 A}{l}}\\
x_{1} & = &   x'_1\sqrt{\frac{C}{ l}}  + x'_2 \sqrt{\frac{C+ 2D}{l}}  \\
x_{2} & = &  x'_1 \sqrt{\frac{C}{ l}}  - x'_2 \sqrt{\frac{C+ 2D}{l}}.
\eea
In the old coordinates the spatial compactification is describe by  the following identification
\be
(x^1, x^2) \sim (x^1+2l, x^2) \sim (x^1, x^2+ 2l) 
\ee 
then in the new coordinates the identifications are given by 
\be
x'^i \sim x'^i + k_{(1)}^i  \sim  x'^i + k_{(2)}^i
\ee
where the lattice vectors are 
\bea
k_{(1)} = \left (\sqrt{\frac{ l^3}{C}}, \sqrt{\frac{ l^3}{C + 2D }} \right )\\
k_{(2)} = \left (\sqrt{\frac{ l^3}{C}}, - \sqrt{\frac{ l^3}{C + 2D }} \right ).
\eea

Thus the spatial manifold has topology of a torus but with a diamond-shaped fundamental domain. Only in the limit $D=0$, which corresponds to suppressed ``neighbor-of-neighbor'' interactions in our lattice mode, the fundamental domain  becomes a square. In higher dimensions the rhombus is replaced with a stretched (along the longest diagonal) hypercube or what is called a hyper-rhombohedron. Note that regardless which longest diagonal is stretched the shortest distances between any two points is same due to compacification (although it obviously depends on the amount of stretching). 

\subsection{Classical Field Theory Solutions}

Now that the field theory is identified as a massive complex scalar field theory on a flat background we can proceed by first analyzing classical solutions of the field equation
\bea
(\partial_\mu \partial^\mu + m^2 ) \varphi =0. \label{eq:Klein-Gordon}
\eea
The only complication is that the spatial fundamental domain of the spatial torus is a hyper-rhombohedron  described by a collection of (generally non-orthogonal)  lattice  vectors $k_{(1)} ...  k_{(N)} $ such that
\be
x^a \sim x^a + \sum_{b} n_b \, k_{(b)}^a \;\;\;\;\;\;\; \forall n_b \in \mathbb{Z}
\ee
for an arbitrary collection of integers $n_b$. It is also convenient to define the dual (or reciprocal) lattice vectors $k^{(1)} ...  k^{(N)} $. Then the most general solution of \eqref{eq:Klein-Gordon} can be written in terms of the dual lattice vectors
\be
\varphi_{cl}(x^\mu) = \sum_{\vec{n} \in \mathbb{Z}^N}  \left( \varphi^-_{\vec n} e^{-i \omega_{\vec n} x^0}+ \varphi^+_{\vec n} e^{i \omega_{\vec n} x^0}\right )\; \exp{\left (- 2\pi i  n_b k^{(b)}_a x^a\right )}  \label{eq:solution}
\ee
where 
\be
\omega_{\vec n} =\sqrt{ \left ( 2\pi  n_b k^{(b)}_a\right )^2  + m^2 }
\ee
and $n_b \in \mathbb{Z}$ is a collection of $N$ integers and summation over repeated $b \in \{1 ... N\}$  is implied. So far the mode coefficients $\varphi^\pm_{\vec n}$ are  arbitrary, but they are to be determined from boundary (i.e. initial and final) conditions. 

It is often important to study transitions from a simple initial state (e.g. $|\psi_{\text{in}}\rangle = |0\rangle$), which in our field theory description can be described with appropriately normalized delta function 
\be
\varphi(x^0=0, x^a) = \det \left | k^{(b)}_a \right | ^{-1}  \prod_{a=1...N} \delta \left (x^a \right ) \label{eq:initial}
\ee
where, without loss of generality, the delta function was placed in the origin. By equating it to \eqref{eq:solution}  at $t=0$ we obtain the following conditions on the mode coefficients, 
\be
 \sum_{\vec{n} \in \mathbb{Z}^N}   \left( \varphi^-_{\vec n} + \varphi^+_{\vec n} \right )\;  \prod_{a=1...N} \delta{\left (\sum_{b=1}^N  (n_b -m_b) k^{(b)}_a \right )}  = 1
\ee
or
\be
\varphi^-_{\vec n} + \varphi^+_{\vec n}   = 1 \;\;\;\;\;\;\; \forall \vec{n} \in \mathbb{Z}^N. \label{eq:mode_condition1}
\ee
Then to solve for transitions to an arbitrary final state $\psi_{\text{out}} \equiv \phi $ all that we have to do is to decompose the final state into Fourier modes 
\be
\varphi(x^0=T, x^a) = \sum_{\vec{n} \in \mathbb{Z}^N} \phi_{\vec n} \; \exp{\left (- 2\pi i \sum_{b=1}^N  n_b k^{(b)}_a x^a\right )} \label{eq:final}
\ee
and match the result to \eqref{eq:solution}  at $t=T$, i.e.
\be
\varphi^-_{\vec n} e^{-i \omega_{\vec n} T}+ \varphi^+_{\vec n} e^{i \omega_{\vec n} T}  = \phi_{\vec n} \;\;\;\;\;\;\; \forall \vec{n} \in \mathbb{Z}^N. \label{eq:mode_condition2}
 \ee
 
We can now solve for mode coefficients using \eqref{eq:mode_condition1} and \eqref{eq:mode_condition2}, i.e.
\be
\varphi^\pm_{\vec n} =\pm  \frac{  \phi_{\vec n}  -  \exp({\mp i \omega_{\vec n} T})  }{2i \sin(\omega_{\vec n} T) }
\ee
which can be substituted into \eqref{eq:solution} to obtain an exact solution 
\be
\varphi_{cl}(x^\mu) = \sum_{\vec{n} \in \mathbb{Z}^N}  \left(  \frac{ \phi_{\vec n} \sin(\omega_{\vec n} x^0)  + \sin(\omega_{\vec n} (T- x^0))  }{\sin(\omega_{\vec n} T) } \right )\; \exp{\left (- 2\pi i \sum_{b=1}^N  n_b k^{(b)}_a x^a\right )}. \label{eq:solution_final}
\ee
This solution (after analytic continuation) can be substituted into Euclidean action to obtain the complexity of state $\psi_{\text{out}} $ with respect to our simple state $\psi_{\text{in}}$.  Of course the hope is that the classical solution (or at least some of them) can be used in constructing efficient quantum circuits which can generate the target state $\psi_{\text{out}} $ from a simple state $\psi_{\text{in}}$. Note that a complication might still arise for small $N$ if we were to bring back the ``global'' constraint which in continuum limit takes the following form:
\be
\int d^Nx\;\; \varphi(t)^* \varphi(t) = 1 \label{eq:normalization}
\ee
for every $t$.  In what follows we will continue to ignore the normalization issue as it is not likely to introduce significant errors in the limit of large $N$. Then the most probable quantum trajectories would be obtained by renormalizing  classical solution \eqref{eq:solution} using the normalization condition \eqref{eq:normalization}.

\section{Quantum Circuits from Classical Solutions}\label{Sec:QuantumCircuits} 

Once the classical solution is obtained we still have to figure out to which quantum circuit it corresponds. To illustrate the procedure (and why it does not always work for the Klein -Gordon theory \eqref{eq:Lagrangian}) we will discretize temporal evolutions of the state vector
\be
|\psi_{(n)} \rangle =  \psi^i_{(n)} |i\rangle 
\ee
and will demand the following mapping of the classical solutions to the evolution of state vector
\be
\psi^i_{(n)} \approx \varphi\left (x^0=t_n, x^a=  k^a_{(b)}i^b/2 \right ) \label{eq:mapping}
\ee
where $i^b$ is the $b$'th digit of integer $i\in \{ 0, 1\}^N$ in base two. Note that the discretization of the time coordinate $0 =t_0 < t_1 < ... < t_n < ... $ in general might not be even, but should be sufficiently fine such that the approximate mapping in Eq. \eqref{eq:mapping} would be satisfied. In fact, if it is true that the computational complexity is given by the Euclidean action \eqref{eq:Euclidean_complexity}, then it makes sense to first re-parametrize the Euclidean time coordinate such that it would increase linearly with complexity along the classical trajectory, i.e.
\be
d\tau_E = L_E(\varphi_{cl}^i, \dot{\varphi}_{cl}^i)  dt_E.
\ee 
Such time coordinate we can call a Euclidean ``proper'' time and then to obtain a Lorentzian ``proper'' time we use the Wick rotation $\tau = -i \tau_E$ prescription. In what follows (and without loss of generality) we will assume that $t$ is already a proper time coordinate (whether it is defined though Euclidean action or directly from the rate of growth of complexity) and thus the time steps are given by $t_n=t_{n-1} + \varepsilon$ where $\varepsilon$ is a sufficiently small but constant number.  

Our next task is to generate a sequence of unitary transformations
 \be
\hat{U}_{(n)} \equiv e^{-i \hat{H}_{(n)}} \label{eq:unitary} 
\ee
such that 
\be
|\psi_{(n)} \rangle = \hat{U}_{(n)} |\psi_{(n-1)} \rangle
\ee
and $\hat{H}_{(n)}$ are Hermitian operators (we shall call Hamiltonians) that can be approximated by the following expression
\bea
\hat{H}_{(n)} &\approx&  i \left ( |\psi_{(n)} \rangle\left (\langle \psi_n | - \langle \psi_{(n-1)} | \right )- (|\psi_{n} \rangle - | \psi_{(n-1)} \rangle \langle \psi_{(n)} | \right )   \notag \\
&=& i \left ( |\psi_{(n)} \rangle  \langle \psi_{(n-1)} | -  | \psi_{(n-1)} \rangle \langle \psi_{(n)} | \right )  \label{eq:hamiltonian0}. 
\eea
These Hamiltonians can be decomposed into the so-called Pauli basis using tensor products of Pauli matrices and identity  (i.e. $\sigma_0=I, \sigma_1=X, \sigma_2=Y$ and $\sigma_3=Z$ ), i.e.
\be
\hat{\sigma}^I = \hat{\sigma}_I \equiv \bigotimes^N_{b=1} \sigma_{I^b} = \sigma_{I^1} \otimes \sigma_{I^2} \otimes ... \otimes \sigma_{I^N} 
\ee
where $I \in \{0,1,2,3\}^N$ is an integer in base four. To emphasize that $\hat{\sigma}^I$ operators are constructed out of tensor products of four matrices we will use capital indices, $I,J,K$, etc. and the summation over repeated indices is always implied unless stated otherwise. Then the Pauli basis decomposition of Hamiltonians is given by
\be
\hat{H}_{(n)} =  H^I_{(n)} \hat{\sigma}_I \label{eq:Pauli_decomposition}
\ee
where 
\be
{H}^I_{(n)}  \equiv 2^{-N} Tr( \hat{H}_{(n)} \hat{\sigma}^I)  \label{eq:Pauli_basis}
\ee
are the Pauli components.  

A working hypothesis (that will turn out to be false in general) is that due to locality of the dual theory used in our construction of the classical solutions the Hamiltonians will be $k$-local and thus could be approximated as a sum of the terms with only one and two- q-bit gates. To make the statement more precise it will be useful to introduce a mapping from integers modulus four to integers modulus two it is convenient to define 
\be
\delta_i[x] \equiv \begin{cases}
1 \;\;\;\text{if}  \;\;\; x = i,\\
0 \;\;\;\text{if}  \;\;\; x \neq i
\end{cases}
\ee
which is nothing but a  Kronecker delta symbol with one of the variables written as a superscript and another variable written in the square brackets. Then if we now define a Pauli weight of the Pauli basis operators as 
\be
w(J) \equiv  \delta_1[J^b]\delta_1[J_b]  +  \delta_2[J^b]\delta_2[J_b]+ \delta_3[J^b]\delta_3[J_b] \label{eq:PauliWeight}
\ee
then it is desired that Pauli components $\hat{H}^I_{(n)}$ with  weight greater than two are suppressed compared to components with weight one or two. This, of course, should be confirmed by direct calculations, which is what we are going to do next.

Consider one of the unitary operators generated by a Hamiltonian 
\bea
\hat{H}_{(n)}= i \left ( \psi_{\text{in},i}  \psi^j_{\text{out}}   -  \psi_{\text{out},j}  \psi^i_{\text{in}}   \right )   |j \rangle \langle i |  \label{eq:hamiltonian} 
\eea
where $\psi_{\text{in}}$ and $\psi_{\text{out}}$ represent the state vector before and after the corresponding unitary operation  is applied. In  \eqref{eq:hamiltonian}  the operator is expressed in computational basis, but we can rewrite it in Pauli basis using  \eqref{eq:Pauli_basis} which gives us
\bea
{H}^I_{(n)}  =  \frac{i}{2^{N}} \left ( \psi_{\text{in},i}  \psi^j_{\text{out}}   -  \psi_{\text{out},j}  \psi^i_{\text{in}}   \right ) \langle i |   \hat{\sigma}^I |j\rangle. 
\eea
Then the Pauli components corresponding to single q-bit gates can be described as (discretized) integrals over the torus. For example, 
\bea
{H}^0_{(n)} & = &   \frac{i}{2^{N}}  \sum_{j\in\{0,1\}^N} \left ( \psi_{\text{in},j}  \psi^j_{\text{out}}    -  \psi^j_{\text{in}}  \psi_{\text{out},j}    \right )\notag\\ & \approx&   \overline { 2 \operatorname{Im} \left(  \varphi(t_{n-1}, x^a) \varphi^*(t_{n}, x^a) \right ) }  \\
{H}^1_{(n)} & =&  \frac{i}{2^{N}}  \sum_{j\in\{0,1\}^N} \left ( \psi_{\text{in},j\oplus1}  \psi^j_{\text{out}}   -  \psi^{j\oplus1}_{\text{in}}  \psi_{\text{out},j}    \right ) \notag\\ & \approx&\overline { 2  \operatorname{Im} \left(\varphi(t_{n-1}, x^a+k^a_{(1)}/2) \varphi^*(t_{n}, x^a) \right ) }   \\
{H}^2_{(n)} & =&  \frac{i}{2^{N}}  \sum_{j\in\{0,1\}^N} i (-1)^{j^1}  \left ( \psi_{\text{in},j\oplus1}  \psi^j_{\text{out}}   - \psi^{j\oplus1}_{\text{in}}  \psi_{\text{out},j}    \right )  \notag\\ & \approx& \;\overline {  2 \exp\left (i  \pi \left ( 2 x^a k^{(1)}_a  +1/2\right )\right )  \operatorname{Im} \left(\varphi\left (t_{n-1}, x^a+k^a_{(1)}/2 \right ) \varphi^*\left (t_{n}, x\right ) \right ) }  \\
{H}^3_{(n)} & =&  \frac{i}{2^{N}}  \sum_{j\in\{0,1\}^N}  (-1)^{j^1}  \left ( \psi_{\text{in},j}  \psi^j_{\text{out}}    -  \psi^j_{\text{in}}  \psi_{\text{out},j}    \right )\notag\\ & \approx& \overline {2 \exp\left (i \pi 2 x^a k^{(1)}_a  \right )  \operatorname{Im} \left( \varphi\left (t_{n-1}, x^a \right ) \varphi^*\left (t_{n}, x^a\right ) \right ) }  
\eea
where $\oplus$ is a bitwise addition modulus two (e.g. $001\oplus001=000, 111\oplus001 = 110, 100\oplus 011 =111$) and the bar denotes spatial average, i.e. 
\be
\overline{f(x)} \equiv  \int \frac{d^N x }{2^{N}} f(x).
\ee
It is also straightforward to obtain the decomposition of the other single q-bit or double q-bit gates. For example,  $H^{21} = Tr \left ( \hat{H} \left ( \hat{X}\otimes \hat{Y} \otimes \hat{I}^{\otimes (N-2)} \right ) \right )$ is given by
\bea
{H}^{21}_{(n)} & =&  \frac{i}{2^{N}}  \sum_{j\in\{0,1\}^N} i (-1)^{j^2}  \left ( \psi_{\text{in},j\oplus21}  \psi^j_{\text{out}}   - \psi^{j\oplus21}_{\text{in}}  \psi_{\text{out},j}    \right )  \\ & \approx& \;\overline { 2 \exp\left (i  \pi \left ( 2 x^a k^{(1)}_a  +1/2\right )\right )  \operatorname{Im} \left(\varphi\left (t_{n-1}, x^a+k^a_{(1)}/2 +k^a_{(2)}/2 \right ) \varphi^*\left (t_{n}, x\right ) \right ) } \notag. \eea
After a bit of algebra one can also show that the most general Pauli component can be written compactly as
\bea
{H}^{J}_{(n)} \approx \;\overline { 2 e^{ i  \pi  \left ( 2 x^a  \left (\delta_2[J_b] +  \delta_3[J_b]\right ) k^{(b)}_a  + \delta_2[J_b]\delta_2[J^b]/2\right )}  \operatorname{Im} \left(\varphi\left (t_{n-1}, x^a+  \left (\delta_1[J^b] +  \delta_2[J^b]\right )  k^a_{(b)}/2 \right ) \varphi^*\left (t_{n}, x\right ) \right ) } \notag \\ \label{eq:components} 
\eea
where summation over repeated indices is implied.  

\section{Abelian-Higgs Dual Theory}\label{Sec:AbelianHiggs}

Note that the Pauli components \eqref{eq:components} are nothing but spatial averages of non-local terms composed of the field operators of  a local scalar field theory described by \eqref{eq:Lagrangian}. Then one might want to argue that the value of the Pauli components of a generic quantum circuit would scale exponentially with distance separating the non-local operators. For example, if $D=0$ and $m$ is sufficiently large, then for Pauli components built by tensor products of only $X$ and $Y$ (and of course $I$) Pauli matrices, we expect that
\be
\langle H^J \rangle_{avg} \propto \exp\left (- m l \sqrt{w(J)} \right ), \label{eq:correlation}
\ee
where the square root in the exponent shows how the distance between field operators scales with Pauli weight \eqref{eq:PauliWeight}. This suggests that the quantum gates simultaneously  acting on a large number of q-bits are exponentially suppressed. This is exactly what we want, but we are not done yet. 

\subsection{The $Z$-problem or Massless Excitations}
The problem is that we cannot say the same about the Pauli components which include tensor products of $Z$  matrices. In fact the tensor products of $Z$ matrices are responsible for the appropriate transformations of phases, but at the level of our scalar field theory the evolution of phases is not constrain to be ``non-relativistic''. The problem with Pauli $Z$ matrices (we call it the $Z$-problem) can be easily seen on the dual theory side by decomposing our complex field using polar coordinates, i.e.
\be
\varphi(x) =  \rho(x) e^{i \theta(x)}
\ee
and then the action \eqref{eq:Lagrangian} can be rewritten as
\be
 {\cal L} = \frac{1}{2}  \partial_\mu \rho \partial^\mu \rho - \frac{1}{2} m^2 \rho^2 + \frac{1}{2} \rho^2 \partial_\mu \theta \partial^\mu \theta. \label{eq:Lagrangian21}
\ee
Roughly speaking the propagation of the field $\theta$ is not suppressed by any mass term and consequently the Pauli components which include multiple $Z$ matrices do not have an exponential suppression which is present for multiple $X$ or $Y$ gates \eqref{eq:correlation}. This becomes even more evident if we consider a Mexican hat potential for the scalar field described by 
\bea
 {\cal L} &= &\frac{1}{2}  \partial_\mu \varphi^* \partial^\mu \varphi + \frac{1}{2} m^2 \varphi^* \varphi - \lambda\left(\varphi^* \varphi\right)^2 \notag  \\
& = & \frac{1}{2}  \partial_\mu \rho \partial^\mu \rho + \frac{1}{2} m^2 \rho^2 - \lambda\rho^4 + \frac{1}{2} \rho^2 \partial_\mu \theta \partial^\mu\theta.
 \label{eq:Lagrangian2}
\eea
Upon spontaneous symmetry breaking the phase degree of freedom  $\theta$ becomes a massless Goldstone excitation which is the source of the $Z$-problem.

Fortunately the problem of massless  excitations  can be solved using Higgs mechanism. What we can do it to promote the global $U(1)$ symmetry of \eqref{eq:Lagrangian2} to a local symmetry using an auxiliary vector field, 
\bea
 {\cal L} & = & \frac{1}{2}D_\mu \varphi^* D^\mu  \varphi + \frac{1}{2} m^2 \varphi^* \varphi - \lambda\left(\varphi^* \varphi\right)^2 -\frac{1}{4}F_{\mu\nu} F^{\mu\nu}   \label{eq:Lagrangian3}
\eea
where 
\be
D_\mu  \equiv \partial_\mu - i q A_\mu
\ee
is a covariant derivative and
\be
F^{\mu\nu} = \partial^{[\mu} A^{\nu]} = \partial^{\mu} A^{\nu}-\partial^{\nu} A^{\mu}
\ee
is the $U(1)$ field tensor. This is a well-known Abelian-Higgs model in which the massless Goldstone mode can be thought to be absorbed by the vector field  $A^\mu$.  The theory  can be analyzed in the so-called unitary gauge where the action takes the following form 
\bea
 {\cal L} & = &  \frac{1}{2}  \partial_\mu \rho \partial^\mu \rho + \frac{1}{2} m^2 \rho^2 -  \lambda \rho^4 - \frac{1}{2} q^2 \rho^2  A_\mu A^\mu- \frac{1}{4}F_{\mu\nu} F^{\mu\nu}     \label{eq:Lagrangian4}.
\eea

In this dual theory the amplitude of our state vector would still be described by $\rho$, but the phase would be describe by the longitudinal component of the vector field $A_\mu$. More precisely we can define an invariant phase $\Theta$ through equation
\be
 \partial_\mu \partial^\mu \Theta = \partial_\mu\left (\partial^\mu \theta - q A^\mu  \right)
\ee
which can be solved for a given solutions of $\partial^\mu \theta - q A^\mu$. Then evolution of the state vector would be described by identification
\be
\psi^i_{(n)} \approx \rho\left (x^0=t_n, x^a=  k^a_{(b)}i^b/2 \right ) r^{i \Theta\left (x^0=t_n, x^a=  k^a_{(b)}i^b/2 \right )} \label{eq:mapping2}
\ee
where both $\rho$ and $\Theta$ are the fields one can solve for. 

Since the vector field $A^\mu$ is massive (with mass  $q m/\sqrt{ 4\lambda}$ in broken phase)  we can adjust the parameters such that correlators of the invariant phase (describe now by $\Theta$)  have a similar exponential suppressions as the correlators of the invariant amplitude (described by $\rho$). Then evolution of the state vector $|\psi\rangle$ would be guaranteed to be confined to only non-relativistic changes, and thus only a small number of $X,Y$ and also $Z$ matricies would be required to reproduce such evolution with quantum circuits. Of course the complications which comes with the proposed modification of the dual theory is that it is no loner quadratic. Moreover the computational task would be a lot more difficult to carry on in practice  since we have to solve the  field theory for all possible initial and final conditions of the transverse modes of the gauge field. 

\subsection{Non-relativistic Limit of Field Theories} 

In a non-relativistic limit we can try to separate the ``rest mass'' contribution using the following ansatz for the scalar field, 
\be
\varphi(x,t) \equiv \phi(x,t) e^{-i m t}.
\ee
Then under assumption that $m \phi \gg  \partial_0 \phi$ we can approximate 
\bea
\partial_0 \varphi^* \partial_0 \varphi -  m^2 \varphi^2 &=& \left ( i m  +  \partial_0 \right ) \phi^*  \left ( - i m + \partial_0 \right ) \phi  -  m^2 \phi^2 \notag\\ & \approx & i m \left ( \phi^* \partial_0 \phi -\phi  \partial_0 \phi^* \right ).
\eea
This can be substituted into Abelian gauge theory Lagrangian  
\bea
 {\cal L} & = & \frac{1}{2}D_\mu \varphi^* D^\mu  \varphi - \frac{1}{2} m^2 \varphi^* \varphi  + V(\varphi^* \varphi ) -\frac{1}{4}F_{\mu\nu} F^{\mu\nu}   \label{eq:Lagrangian3}
\eea
to yield (after integration by parts and setting without loss of generality $m=1$) a non-relativistic theory described by 
\bea
 {\cal L} & = &  i \phi^* D_0 \phi + \frac{1}{2} D_a \phi^*D^a \phi  - V(\phi^* \phi )  -\frac{1}{4}F_{\mu\nu} F^{\mu\nu}    \label{eq:Lagrangian5}.
\eea
where $V(\phi^* \phi )$ is an interaction term which may or may not be zero.  Upon variation we obtain a (in general non-linear) Schrodinger equation
\be
i \frac{\partial \phi}{\partial t} = \left ( i q A_0 + D_a D^a   \right) \phi + V(\phi^* \phi)  \label{eq:Schrodinger}
\ee
and Maxwell's equation
\be
\partial_\mu F^{\mu\nu} =  J^\mu
\ee
where 
\bea
J^0 &\equiv& \phi^* \phi \\
J^a &\equiv& \frac{i}{2} \left ( \phi^* D^a \phi - \phi D^a \phi^*  \right )
\eea
are the conserved (non-relativistic) charge and current densities. Then the conserved charge density 
\be
J^0 = \phi^* \phi = \rho^2
\ee
is exactly what we want for the evolution of scalar field  to describe the state vector (upon identification \eqref{eq:mapping2}), whose normalization condition $\psi_{i}(t) \psi^i(t) =1$ would be automatically satisfied at all times if it is satisfied at the initial time. Moreover, whenever the interactions are suppressed  (i.e.  $V(\varphi^* \varphi ) \approx 0$)  the orthonormal states would remain orthonormal throughout evolution. This property is essential for the applications of our methods to the problem of construction of arbitrary unitary operators that we shall discuss very briefly. 

Consider an arbitrary unitary operator $\hat{U}$. Our task is to create a quantum circuit that would transfer all of the (orthonormal) coordinate basis initial states $|i\rangle$ into final states,
\be
|\psi_{[i]}\rangle = \hat{U} |i \rangle.
\ee
So far we have leaned how to construct a quantum circuit that would approximate a unitary evolution of a single initial state (e.g. $|0\rangle$) to a single final state (e.g. $| \psi_{[i]} \rangle $) with a discrete set of unitary operators
 \be
\hat{U}_{(n)} \equiv e^{-i \hat{H}_{(n)}}
\ee
where $ \hat{H}_{(n)}$ was given by \eqref{eq:hamiltonian0}. But now we have to ensure that all of the orthonormal states are correctly evolved by our discrete sequence of unitary operators, $\hat{U}_{(n)}$. For that we set the Hamiltonian operators to be given by a sum of the terms determined from the evolutions of individual state vectors, i.e.
\bea
\hat{H}_{(n)} &\approx& i \sum_j \left ( |\psi_{[j],(n)} \rangle  \langle \psi_{[j],(n-1)} | -  | \psi_{[j],(n-1)} \rangle \langle \psi_{[j],(n)} | \right )  \label{eq:hamiltonian2}. 
\eea
(Note that we slightly  abuse the notations and use the first index in square brackets  to enumerate different initial vectors and the second index in round brackets to enumerate its (discrete) time evolution.) Of course, now the task is a lot more general and a lot more difficult to carry on in practice  since we have to solve the quantum field theory for $2^N$ transitions from different initial (determined by $|i\rangle$) to different final states (determined by $\hat{U} |i\rangle$), but this is not an exponentially (in $2^N$) difficult task and thus does not possess a huge problem. 

There is however a potentially more serious problem. Whenever the interaction term  $V(\varphi^* \varphi )$ in Lagrangian \eqref{eq:Lagrangian5} is not negligible, the orthonormal states would still remain normal (in non-relativistic limit), but might not remain orthogonal. As a result the quantum circuit (built from the Hamiltonians in \eqref{eq:hamiltonian2}) might not describe the unitary evolution $\hat{U}$ that we would like to approximate.

\section{Solving The Quantum Maze}\label{Sec:QuantumMaze} 

Now that the dual theory description of the quantum maze problem is constructed we can discuss algorithms for finding the shortest quantum circuit which connects an arbitrary pair of states. The goal would be to estimate the classical computational complexity of such algorithms (not to confuse with the quantum computational complexity of shortest quantum circuit).  Our model depends on at least nine parameters: $A, B$ (or $m^2$ given by \eqref{eq:mass}),  $C, D, l, T, \varepsilon, \lambda, q$  that we can try to fix. The metric parameters $A$ and $C$ can both be set to $1$ without loss of generality and the temporal parameter $T$ would  scale with the computational complexity of a given computational task and the time step $\varepsilon$ must remain sufficiently small (generically $2^N$ times smaller compared to $T$) so that an approximation \eqref{eq:hamiltonian0} makes sense. Since we only want to suppress contributions of the three or more q-bit gates, but not the two q-bit gates we can adjust $D$ so that the expected Pauli coefficients with Pauli weights $1$ and $2$ are of the same order. Moreover, the exponential suppression of correctors (discussed in Sec. \ref{Sec:AbelianHiggs})  should be at the scale of lattice spacing $l$ which lives us with at least three parameters: mass scale $m$, coupling constant $\lambda$, and charge $q$. All these parameters are to be determined ``numerically'' and it would be interesting to see how they scale with the number of q-bits $N$. We leave all these questions for future work, and will instead discuss an algorithm for solving quantum maze and its computational complexity.

The largest (and also typical) quantum computational complexity of transitioning from a simple state to a given target state scales as $2^N$. Since we want to estimate the complexity of our algorithm in the worst possible case we can a priori set $\varepsilon=1$ and $T=2^N$. Then all that we want to estimate is how difficult it is to figure out what should be the first quantum gate. If the answer is polynomial in $2^N$ then multiplying by an additional factor of $2^N$ hardly changes anything and the quantum maze problem would be solvable in a polynomial in $2^N$ time. The initial and final states for $\varphi$ are fixed but the main problem is that there is still a freedom of choosing transverse modes for the gauge field. Since the Abelian-Higgs theory is not exactly solvable we cannot study a continuum of possible choices for the gauge field and thus the field must be discretized. This makes the task of figuring out what would be the first gate exponentially hard in $2^N$ and so the entire algorithm is exponentially difficult to implement.  Although we managed to fix the $Z$-problem using Higgs mechanism we are back to the fact that in general it would be exponentially difficult to find a solution using the Abelian-Higgs dual theory. 

This is a good place where the dual theory approach can be compared to a geometric approach developed in  Refs. \cite{Nielsen, Nielsen2, Nielsen3}. There the authors describe the problem of finding the shortest  $N$ q-bit quantum circuit in terms of geodesic distances in $2^N$ dimensional space. In our approach the problem is formulated in terms of semiclassical theories of quantum fields in $N+1$ dimensional space-time with geometrically flat, but topologically compact spatial slices. Unfortunately both approaches, i.e. the geometric and the field theoretic, lead to the same conclusion - the quantum maze problem is an exponentially hard to solve. Does it mean that we have exhausted all of the possibilities? Of course not.  

So far we have only considered field theories on flat spacetime although the spatial fundamental domain was an $N$-dimensional torus. The computationally complex trajectories in the field theory configuration space we punished using the mass terms. But one might ask if we can accomplish the same by making the geometry curved and/or the topology less trivial. This is along the lines of reasoning used in Ref. \cite{Nielsen} with an important difference that we are talking about $N+1$-dimensional geometry instead of $2^N$ dimensional geometry. However, we must be careful not to introduce any additional degrees of freedom and so we shall insist that the metric tensor $g_{\mu\nu}$ is not dynamical, but the dual filed theory is once again a simple Klein-Gordon theory. 

Then there are at least three possible geometric and/or topological solutions that we can describe without going into details. If we allow the spatial geometry to curve all that we need to do is to curve it in such a way that solutions for Klein-Gordon fields stay away from all diagonals connecting vertices of the lattice sites $i$ and $j$ that have Hamming distance $h(i,j) > 2$. One possibility is to have only positive spatial curvature along all paths connecting vertices $i$ and $j$  with  Hamming distance $h(i,j) =1 $ or $2$ and negative curvature along all path connecting vertices with Hamming distance $h(i,j) > 2$. The author is not aware of any theorem that would say that it is impossible, but is also not yet ready to write down the metric tensor and so this will be left for future work.

Another possibility is to cut-out the interior from every lattice cube. (Altogether there are $2^N$ lattice cubes because the fundamental domain of the $N$-torus has volume $(2l)^N$ which is $2^N$ times larger than the volume of a single lattice cube whose volume is $l^N$.) After cutting out a ``significant'' portion (i.e. of size $\lesssim l^N$) of each lattice cube the manifold is no longer compact, and there are least two possibilities. We can either impose reflecting boundary conditions along all cuts or we can try to make identifications that will make the manifold compact again. (In both cases we would also have to smooth out the geometry near corners of these cuts to avoid caustics). Note that the identifications cannot be arbitrary as we still want the geodesic distances between all points with large Hamming distance to be large. Once again the author is not aware of any theorem that says that it is impossible, but the exact construction of such manifolds will be left for future work.

\section{Summary}\label{Sec:Summary}

In this paper we considered a problem of finding the shortest quantum circuit consisting of only one- and two- q-bit gates that would transfer an arbitrary $N$ q-bit initial state to an arbitrary $N$ q-bit  final state. We called it the ``quantum maze'' problem and argued that it is relevant to both quantum computation and quantum gravity. Until now, the only systematic and generic treatment of the problem was given from a purely geometric prospective \cite{Nielsen} where the  quantum maze problem (more precisely a related problem) was shown to be equivalent to the problem of finding the shortest geodesics in $2^N$ dimensional curved space. Instead of focusing on the geometric ideas we made an attempt to map the quantum maze problem to a dual field theory problem (generically on a curved background) that we might know how to solve.

In particular it was first argued that the quantum maze problem is equivalent to the problem of finding semiclassical trajectories in some lattice field theories (the dual theories) on an $N+1$ dimensional space-time with geometrically flat, but topologically compact spatial slices. The spatial fundamental domain was an $N$ dimensional hyper-rhombohedron, and the temporal direction described transitions from an arbitrary initial state to an arbitrary target state.  We then considered a simple complex Klein-Gordon field theory in $N+1$ dimensional space-time  with compact spatial domain and argued that such dual theory can only be used to study the shortest quantum circuits which do not involve generators composed of tensor products of multiple Pauli $Z$ matrices.  One can call such quantum circuits and the corresponding target states  $Z$-simple. However, the $Z$-simple quantum circuits (or states) are not generic which is what we called the $Z$-problem.  On the dual field theory side the $Z$-problem corresponds to the well known problem of massless excitations of the phase or the Goldstone mode. 

To fix the $Z$-problem on the quantum computation side we first proposed to used a Higgs mechanism on the dual theory side. The simplest dual theory  which does not suffer from the massless excitation (or from the $Z$-problem) is an Abelian-Higgs model which we argue can be used for finding the shortest quantum circuits. Since every trajectory of the field theory is mapped directly to a quantum circuit,  the shortest quantum circuits (consisting of a finite number of gates) were identified with semiclassical filed theory trajectories.  Although the $Z$-problem is fixed with Higgs mechanism another problem was introduced due to addition of the new degrees of freedom in the Abelian-Higgs dual theory (i.e. transverse modes of gauge field). It turned out that because of these new degrees of freedom the algorithmic complexity to actually solve the quantum maze problem remained exponentially hard even if in the Abelian-Higgs dual description is employed. 

Then we argued that if our main task is to come up with an algorithm which is sub-exponential in time we must leave the field content minimal, i.e. complex Klein-Gordon field, and change the geometry and/or topology of our $N+1$ dimensional spacetime. We discussed three possible modification of the spatial geometry and/or topology that looked rather promising. The first involved leaving topology as is, but adding negative curvature along trajectories that we wanted to punish (i.e. connecting vertices with Hamming distance $3$ or more) and positive curvature along trajectories which we want to reward (i.e. connecting vertices with Hamming distance $1$ or $2$). The second and third modification to the manifold involved cutting certain regions from the inside of each of the $2^N$ lattice cubes and then imposing either reflecting boundary conditions, or making identifications so that the manifold remains compact.  Detailed calculations base on these geometric and/or topological ideas were left for future work.

{\it Acknowledgments.} The author is grateful to McCoy Becker, Adam Brown, Yi-Zen Chu, Arash Fereidouni, Andrey Grabovsky, Mudit Jain, Lenny Susskind and Brian Swingle for very useful discussions and comments on the manuscript. The work was supported in part by Templeton Foundation and Foundational Questions Institute (FQXi).


\begin{thebibliography}{10}

\bibitem{book}
M. Nielsen and I. Chuang, 
``Quantum Computation and Quantum Information, ''
Cambridge University Press, 2010.

\bibitem{Brown:2015bva} 
  A.~R.~Brown, D.~A.~Roberts, L.~Susskind, B.~Swingle and Y.~Zhao,
  ``Complexity Equals Action,''
  arXiv:1509.07876 [hep-th].
 
  
\bibitem{Harlow} 
  D. Harlow and P. Hayden, 
  ``Quantum computation vs. firewalls,''
   Journal of High Energy Physics 6 (June, 2013) 85, 1301.4504.
  
  
\bibitem{Susskind1}
L. Susskind, 
``Computational Complexity and Black Hole Horizons,''
arXiv:1402.5674


\bibitem{Susskind2}
L. Susskind, 
``Addendum to Computational Complexity and Black Hole Horizons,''
arXiv:1403.5695.


\bibitem{Firewall}
A. Almheiri, D. Marolf, J. Polchinski, and J. Sully, 
``Black Holes: Complementarity or Firewalls?,''
 arXiv:1207.3123.
 
 
\bibitem{Swingle}
B.~Swingle,
  ``Entanglement Renormalization and Holography,''
  Phys.\ Rev.\ D {\bf 86}, 065007 (2012)
  
  
 \bibitem{Dong} 
  A.~Almheiri, X.~Dong and D.~Harlow,
 ``Bulk Locality and Quantum Error Correction in AdS/CFT,''
  JHEP {\bf 1504}, 163 (2015)
 
 \bibitem{Pastawski} 
  F.~Pastawski, B.~Yoshida, D.~Harlow and J.~Preskill,
  ``Holographic quantum error-correcting codes: Toy models for the bulk/boundary correspondence,''
  JHEP {\bf 1506}, 149 (2015)
  
  
  
  
  \bibitem{Czech} 
  B.~Czech, L.~Lamprou, S.~McCandlish and J.~Sully,
 ``Integral Geometry and Holography,''
  JHEP {\bf 1510}, 175 (2015)
  
  
 \bibitem{Brown:2015lvg} 
  A.~Brown, D.~A.~Roberts, L.~Susskind, B.~Swingle and Y.~Zhao,
  ``Complexity, Action, and Black Holes,''
  arXiv:1512.04993 [hep-th].

 
  
\bibitem{Nielsen}
Mark R. Dowling, Michael A. Nielsen,
``The geometry of quantum computation'',
arXiv:quant-ph/0701004

\bibitem{Nielsen2}
M. A. Nielsen, M. Dowling, M. Gu, A. Doherty,
``Quantum Computation as Geometry'',
 Science 311, 1133 (2006)
  
 \bibitem{Nielsen3}
Michael A. Nielsen, Mark R. Dowling, Mile Gu, and Andrew C. Doherty, 
``Optimal control, geometry, and quantum computing'',
Phys. Rev. A 73, 062323 

  
\end{thebibliography}
\end{document}